\documentclass[longtitle,authoryear,12pt]{elsarticle}
\usepackage{graphicx}
\usepackage{subfig}
\usepackage{wrapfig}
\usepackage{tikz}
\usetikzlibrary{positioning}
\tikzset{%
  cont state/.style={circle, draw=black!, thick, inner sep=0pt,
                     node distance=3mm, minimum size=8mm,
                     font=\small},
  disc state/.style={rectangle, draw=black!, thick,
                     inner sep=0pt, node distance=3mm,
                     minimum size=8mm, font=\small},
  meas/.style={circle, draw=black!, fill=black!20,  thick, inner
               sep=0pt, node distance=3mm, minimum size=8mm,
               font=\small},
  empty/.style={node distance=5mm}}

\usepackage{xspace}
\newcommand*\ie{\emph{i.e.}\xspace}
\newcommand*\eg{\emph{e.g.}\xspace}

\usepackage{color}
\usepackage{amsmath}
\usepackage{amsfonts}
\usepackage{amssymb}
\usepackage{mathtools}
\usepackage[english]{babel}
\usepackage[latin1]{inputenc}
\usepackage{algorithm}
\usepackage{enumitem}
\newtheorem{rem}{Remark}
\usepackage{stfloats} 
\DeclareMathOperator*\Var{Var}

\graphicspath{{./}{./figures/}}
\begin{document}
\thispagestyle{empty}

\begin{frontmatter}


  \title{Online inference in Markov modulated nonlinear dynamic
    systems: a Rao-Blackwellized particle filtering approach}


\author[LiU]{Saikat Saha}
\author[LiU,FOI]{Gustaf Hendeby}

\address[LiU]{%
  Dept.\ Electrical Engineering,
  Div.\ Automatic Control,\\
  Linköping University,
  SE 581-83 Linköping, Sweden,\\
  e-mail: \{saha,hendeby\}@isy.liu.se}
\address[FOI]{%
  Div.\ Sensor Informatics,
  Dept.\ Sensor \& EW Systems,\\
  Swedish Defence Research Agency (FOI),
  SE-581 11 Linköping, Sweden}

\begin{abstract}
  The Markov modulated (switching) state space is an important model
  para\-digm in applied statistics. In this article, we specifically
  consider Markov modulated nonlinear state-space models and address
  the online Bayesian inference problem for such models. In
  particular, we propose a new Rao-Blackwellized particle filter for
  the inference task which is our main contribution here.
  % by marginalizing out the switching (regime) variable analytically.
  The detailed descriptions including an algorithmic summary are
  subsequently presented.
\end{abstract}


\end{frontmatter}
% \linenumbers
\thispagestyle{empty}

\section{Introduction}

In many practical applications of applied science, engineering and
econometrics, one often deals with nonlinear dynamic systems involving
both a continuous value target state and a discrete value regime
variable. Such descriptions imply that the system can switch between
different nonlinear dynamic regimes, where the parameters of each
regime is governed by the corresponding regime variable. The different
regimes can possibly be described in terms of different stochastic
processes. The regime variable also evolves dynamically according to a
finite state Markov chain. Both the target state and regime variable
are latent and are related to the noisy observations. This model
paradigm is often referred to as a \emph{Markov regime switching}
(MRS) state space, sometimes with other monikers like jump Markov,
Markov modulated or hybrid dynamic system. Due to its modeling
flexibility, MRS is very popular in different disciplines and as such,
has been successfully used in diverse areas like econometrics,
operations research, control theory, and signal processing, population
dynamics, and machine learning among others \citep{Fruhwirth07,
  Risticbook04, Luo07, mamon2007hidden, barberBRML2012}. However, most
of the studies have focused on a special case where each individual
regime follows a linear Gaussian state-space model. This special case
is known as the \emph{jump Markov linear system} (JMLS). Nonetheless,
for many practical applications of interest, including econometrics
\citep{carvalhoHL07}, signal processing \citep{AndrieuMDAD03}, target
tracking and localization \citep{Hans05} are some examples, the
individual regimes follows nonlinear dynamics, possibly driven by
non-Gaussian processes. Such a system is referred to as a \emph{Markov
  modulated nonlinear dynamic system} (MmNDS) or a \emph{jump Markov
  nonlinear system} (JMNS). Compared to JMLS, this class of problem is
less well studied. Hence we consider the state inference problem for
MmNDS.

For certain models, part of the state space may be (conditionally)
tractable. It is then sufficient to employ a \emph{particle filter}
(PF) for the remaining intractable part of the state space. By
exploiting such analytical substructure, the Monte Carlo based
estimation is then confined to a space of lower
dimension. Consequently, the estimate obtained is often better and
never worse than the estimate provided by the PF targeting the full
state space. This efficiency arises due to the implication of the
well-known \emph{Rao-Blackwell} estimator (see the Appendix). For this
reason, the resulting method is popularly known as
\emph{Rao-Blackwellized particle filtering} (RBPF)
\citep{chen:liu:00:mixture, Doucet2000a, Chopin04, Schon05, Hendeby10,
  SahaOGS09}.

In this article, we address the online inference problem for MmNDS
using PF. Particularly, we propose a new RBPF framework using the
conditionally analytical substructure of the regime indicator
variable. To the best of our knowledge, this RBPF framework has not
yet been exploited.

The organization of the article is as follows. In
Section~\ref{PF:intro}, we provide a brief but necessary PF
background. This is followed by the problem statement in
Section~\ref{prob_state}, where we first describe the model and then
pose the inference objective. In Section~\ref{RBPF:desc}, the
derivations for the new RBPF scheme are outlined for one complete
cycle and an algorithm is also presented. In Section~\ref{RBPF_comp},
we provide comparisons to other similar models. Finally, we concluded
in Section~\ref{conclusion}.

\section{Brief background on Particle filter (PF)}
\label{PF:intro}

Consider the following discrete time general state-space model
relating the latent state $x_{k}$ to the observation $y_{k}$ as
\begin{subequations}
  \begin{align}
    x_k
    &= f(x_{k-1}, w_{k-1}),\\ %
    y_k
    &=  h(x_{k}, e_k) ,
  \end{align}
\end{subequations}
where $f(x_{k-1}, w_{k-1})$ describes how the state propagates driven
by the process noise $w_{k-1}$, and $h(x_k, e_t)$ describes how the
measurements relates to the state and how the measurement is affected
by noise, $e_k$.  This model can also be expressed with a
probabilistic model
\begin{subequations}
  \label{eq:model}
  \begin{align}
    \label{transition}
    x_{k}
    &\sim  p(x_{k} |x_{k-1}),\\ %
    \label{Obsvn}
    y_{k}
    &\sim  p(y_{k} |x_{k}),
  \end{align}
\end{subequations}
% where $\theta$ $\in \Theta \subset \mathbb{R}^{m}$ is the static
% parameter vector.
where $p(x_{k} |x_{k-1})$ and $p(y_{k} |x_{k})$ are the corresponding
state transition and observation likelihood densities, which are here
assumed to be known. Given this model, the density for the initial
state (\ie, $p(x_0)$) and the stream of observations
$y_{0:k}\triangleq \{y_0,y_1,\dots,y_k\}$ up to time $k$, the
inference objective is to optimally estimate the sequence of posterior
densities $p(x_{0:k} |y_{1:k})$, and typically their marginals
$p(x_{k}|y_{1:k})$, over time.  The above posteriors are in general
intractable but can be approximated using PF to arbitrary accuracy. In
PF, the posterior distribution associated with the density $p(x_{0:k}
|y_{1:k})$ is approximated by an empirical distribution induced by a
set of $N$ weighted particles (samples) as
\begin{equation}
  \widehat{P}_{N}(dx_{0:k}|y_{1:k})
  = \sum_{i=1}^{N}\widetilde{w}_{k}^{(i)}\delta_{x_{0:k}^{(i)}}(dx_{0:k}),
\end{equation}
where $\delta_{x_{0:k}^{(i)}}(A)$ is a Dirac measure for a given
$x_{0:k}^{(i)}$ and a measurable set~$A$, and
$\widetilde{w}_{k}^{(i)}$ is the associated weight attached to each
particle $x_{0:k}^{(i)}$, such that
\mbox{$\sum_{i=1}^{N}\widetilde{w}_{k}^{(i)}=1$}.  Given this PF
output, one can approximate the marginal distribution associated with
$p(x_{k}|y_{1:k})$ as
\begin{gather}
  \widehat{P}_{N}(dx_{k}|y_{1:k})
  = \sum_{i=1}^{N}\widetilde{w}_{k}^{(i)}\delta_{x_{k}^{(i)}}(dx_{k}), %
\shortintertext{and expectations of the form} %
  \label{mean_es}
  I(g_k)
  = \int g_k\left(x_{0:k}\right)p\left(x_{0:k}|y_{1:k}\right)dx_{0:k}
\end{gather}
as
\begin{subequations}
  \begin{align}
    \widehat{I}_{N}(g_k)
    &= \int g_k\left(x_{0:k}\right)\widehat{P}_{N}(dx_{0:k}|y_{1:k})\\ %
    \label{MC_INT}
    & \approx \sum_{i=1}^{N}\widetilde{w}_{k}^{(i)}g_k(x_{0:k}^{(i)}).
  \end{align}
\end{subequations}
Even though the distribution $\widehat{P}_{N}(dx_{0:k}|y_{1:k})$ does
not admit a well defined density with respect to the Lebesgue measure,
the density $p(x_{0:k} |y_{1:k})$ is conventionally represented as
\begin{equation}
  \label{PF:density}
  \widehat{p}_{N}(x_{0:k}|y_{1:k})
  = \sum_{i=1}^{N}\widetilde{w}_{k}^{(i)}\delta(x_{0:k}-{x_{0:k}^{(i)}}),
\end{equation}
where $\delta(\cdot)$ is the Dirac-delta function. The notation used
in \eqref{PF:density} is not mathematically rigorous; however, it is
intuitively easier to follow than the stringent measure theoretic
notations. This is especially useful if we are not concerned with
theoretical convergence studies.

Now suppose at time $k-1$, we have a weighted particle approximation
of the posterior $p(x_{0:k-1}|y_{1:k-1})$ as
$\widehat{P}_{N}(dx_{0:k-1}|y_{1:k-1})
=\sum_{i=1}^{N}\widetilde{w}_{k-1}^{(i)}\delta_{x_{0:k-1}^{(i)}}(dx_{0:k-1})$. With
the arrival of a new measurement $y_{k}$, we wish to approximate
$p(x_{0:k}|y_{1:k})$ with a new set of samples.  The particles are
propagated to time $k$ by sampling a new state $x_{k}^{(i)}$ from a
proposal kernel $\pi(x_{k}|x_{0:k-1}^{(i)},y_{1:k})$ and setting
$x_{0:k}^{(i)} \triangleq \left(x_{0:k-1}^{(i)}, x_k^{(i)}\right)$.
Since we have
\begin{equation}
  p(x_{0:k}|y_{1:k}) \propto p(y_{k}|x_{0:k},y_{1:k-1})\,
  p(x_{k}|x_{0:k-1},y_{1:k-1})\,p(x_{0:k-1}|y_{1:k-1})
\end{equation}
and using the Markovian property \eqref{eq:model}, the corresponding
weights of the particles are obtained as
\begin{align}
  \label{WT_rec}
  w_{k}^{(i)}
  &\propto \widetilde{w}_{k-1}^{(i)}
  \frac{p(y_{k}|x_{k}^{(i)})p(x_{k}^{(i)}|x_{k-1}^{(i)})}
  {\pi(x_{k}^{(i)}|x_{0:k-1}^{(i)},y_{1:k})}\\ %
  \widetilde{w}_{k}^{(i)}
  &= \frac{{w}_{k}^{(i)}}{\sum_{j=1}^{N}{w}_{k}^{(j)}}.
\end{align}
%The weights are then normalized as $\widetilde{w}_{k}^{(i)} = \frac{{w}_{k}^{(i)}}{\sum_{i=1}^{N}{w}_{k}^{(i)}}$.

To avoid carrying trajectories with small weights and to concentrate
upon the ones with large weights, the particles need to be resampled
regularly.  When resampling, new particles are sampled with
replacement from the old ones with the probabilities
$\{\widetilde{w}_{k}^{(i)}\}_{i=1}^N$.  The effective sample size
$N_{\text{eff}}$, a measure of how many particles that actually
contributes to the approximation of the distribution, is often used to
decide when to resample.  When $N_{\text{eff}}$ drops below a
specified threshold, resampling is performed.  For a more general
introduction to PF, refer to \cite{Doucet:Johansen11}.

\section{Problem Statement}
\label{prob_state}

In this section, we first provide a description of the model and
subsequently pose the estimation objectives.

\subsection{Model description:}

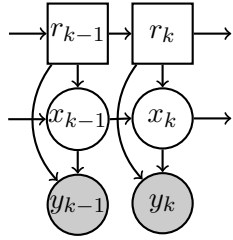
\begin{figure}
  \centering
  \begin{tikzpicture}
  \node[disc state] (r2) {$r_{k-1}$};
  \node[disc state] (r3) [right = of r2] {$r_{k}$};
  \node[empty] (r1) [left = of r2] {};
  \node[empty] (rN) [right = of r3] {};

  \node[cont state] (x2) [below = of r2] {$x_{k-1}$};
  \node[cont state] (x3) [below = of r3] {$x_k$};
  \node[empty] (x1) [left = of x2] {};
  \node[empty] (xN) [right = of x3] {};

  \node[meas] (y2) [below = of x2] {$y_{k-1}$};
  \node[meas] (y3) [below = of x3] {$y_k$};

  % Draw the lines
  \draw[thick] [->] (r1.east) -- (r2.west);
  \draw[thick] [->] (r2.east) -- (r3.west);
  \draw[thick] [->] (r3.east) -- (rN.west);

  \draw[thick] [->] (r2) -- (x2);
  \draw[thick] [->] (r3) -- (x3);

  \draw[thick] [->] (x1.east) -- (x2.west);
  \draw[thick] [->] (x2.east) -- (x3.west);
  \draw[thick] [->] (x3.east) -- (xN.west);

  \draw[thick] [->] (x2.south) -- (y2.north);
  \draw[thick] [->] (x3.south) -- (y3.north);

  % Draw the curves from r_i to y_i
  \draw[thick] [->] (r2) to [bend right=40] (y2);
  \draw[thick] [->] (r3) to [bend right=40] (y3);
\end{tikzpicture}

%%% Local Variables:
%%% mode: latex
%%% mode: TeX-PDF
%%% TeX-master: "../bare_jrnl.tex"
%%% End:
  \caption{Graphical representation of a Markov modulated nonlinear
    dynamic systems.}
  \label{fig:MSNDM}
\end{figure}

Consider the following (hybrid) nonlinear state-space model evolving
according to
\begin{subequations}
  \label{eq:hybrid-model}
  \begin{align}
    \label{mode_dyn}
    \Pi(r_{k}|r_{k-1}),\\       %
    \label{reg_dyn}
    p_{\theta_{r_{k}}}(x_{k}|x_{k-1},r_{k}),\\ %
    p_{\theta_{r_{k}}}(y_{k}|x_{k},r_{k}),
  \end{align}
\end{subequations}
where $r_{k}\in S \triangleq \{1,2,\dots,s\}$, is a (discrete) regime
indicator variable with finite number of regimes (\ie, categorical
variable), $x_{k}\in \mathbb{R}^{n_x}$ is the (continuous) state
variable. As the system can switch between different dynamic regimes,
for a given regime variable $l \in S$, the corresponding dynamic
regime can be characterized by a set of parameters $\theta_{l}$. Both
$x_{k}$ and $r_{k}$ are latent variables, which are related to the
measurement $y_{k}\in \mathbb{R}^{n_y}$. The time behavior of the
regime variable $r_{k}$ is commonly modeled by a homogeneous
(time-invariant) first order Markov chain with \emph{transition
  probability matrix} (TPM) $\Pi =\left[ \pi_{ij} \right]_{ij}$ as
\begin{subequations}
  \label{eq:mode-prob}
  \begin{gather}
    \pi_{ij}
    \triangleq  \mathbb{P}(r_k = j | r_{k-1} = i) \ \ \ (i,j\in S),\\ %
    \label{mode_prob}
    \pi_{ij} \ge 0; \qquad \sum_{j=1}^{s}{\pi_{ij}}= 1,
  \end{gather}
\end{subequations}
This model is represented graphically in Figure~\ref{fig:MSNDM}.
%\TODO{provide GRAPHICAL REPRESENTATION }
%\subsubsection{\TODO{Example 1}}
%\subsubsection{Example 2}
%\subsubsection{Example 3}
We also present below the following examples illustrating some real
life applications where the above model is used.

\textbf{Example 1}: Consider the Markov switching stochastic
volatility model \citep{carvalhoHL07}, where $x_{k}$ is the latent
time varying log-volatility, $y_{k}$ is the observed value of daily
return of stock price or index. The regime variable $r_{k}$ is modeled
as a \emph{$K$-state} first order Markov process. The model is further
specified as
\begin{subequations}
  \begin{align}
    p_{\theta_{r_{k}}}(x_{k}|x_{k-1},r_{k})
    &= \mathcal{N}(\alpha_{r_k}+\phi x_{k-1},\sigma^{2}), \\ %
    p(y_{k}|x_{k},r_{k})
    &= \mathcal{N}(0,e^{{x_k /2}}),
  \end{align}
\end{subequations}
where the parameter vector is given by $\theta_{r_{k}} \triangleq
\{\alpha_{r_k}, \phi, \sigma \}$.

\textbf{Example 2}: Consider an altitude based terrain navigation
framework \citep{Schon05}.  To keep the description simple, assume
that an aircraft is traveling in an one dimensional space (\eg, on a
manifold). The aircraft is assumed to follow a constant velocity
model. The state-space model is given as
\begin{subequations}
  \begin{align}
    \label{eq:ternav:x}
    x_{k+1}
    &=
    \begin{pmatrix}
      1 & T\\
      0 & 1\\
    \end{pmatrix} x_k
    +
    \begin{pmatrix}
      \frac{1}{2}T^2\\ T
    \end{pmatrix} w_k\\ %
    y_k
    &= h(x_k)
    + e_k(r_k) ,
  \end{align}
\end{subequations}
where $T$ is the sampling period, $w_k$ and $e_{k}(\cdot)$ are the
process and the measurement noise, respectively, commonly assumed to
be individually independent and also independent of each other. The
aircraft latent state $x_k$ consists of position and velocity. The
observation $y_k$ denotes the terrain altitude measured by the
aircraft at time $k$. This is obtained by deducting the height
measurement of the ground looking (on board) radar from the known
altitude of the aircraft (obtained using an altimeter). The function
$h(x_k)$ relates the terrain altitude to position $x_k$ in the form of
a digital terrain database. As the same height can corresponds to
different locations, $h(\cdot)$ is highly nonlinear.

The distribution of $w_k$ is typically modeled as Gaussian. As radar
reflections can come from the ground as well as from the tree canopy,
typically the observation noise $e_k$ is modeled as a (bimodal) two
component Gaussian mixture. The regime variable $r_k$ indicates the
corresponding mixture component. The sufficient statistics (\ie, mean
and variance) of each component can be specified by the regime
dependent parameters $\theta_{r_{k}}$. The dynamics of $r_k$ is
modeled as two state first order homogeneous Markov process.

%  The following measurement noise has been used
%\begin{equation*}
%  e_t
%  \sim \tfrac{3}{4}\Npdf(0, 9)
%  + \tfrac{1}{4}\Npdf(12, 36) ,
%\end{equation*}
% The Gaussian mixture
%represents a typical radar response from height measurements over a
%forest.  The noise types and levels are typical for the application,
%\cite{Nordlund:02a}.  See also Example~\ref{ex:NGausspdf} for an
%illustration of typical radar noise.
%
%
%-----------------------------------
%

\subsection{Inference objective:}

For the model described
by~\eqref{eq:hybrid-model}--\eqref{eq:mode-prob}, given the densities
for the initial state $\{r_0,x_0\}$ and the measurements up to a time
$k$, our interest lies in estimating sequentially the latent states
$\{r_k,x_k\}$. More precisely, for the statistical inference purpose,
we target the series of filtering distributions
$\mathbb{P}(r_{k}|y_{1:k})$ and $p(x_{k}|y_{1:k})$ recursively over
time.  However, the above posteriors are in general, computationally
intractable. Given this intractability, PF is a suitable candidate for
this approximate (real time) inference task. Interestingly, however,
we note that conditioned on the sequence $x_{1:k}$, $r_k$ follows a
finite state \emph{hidden Markov model} (HMM), implying that
$\mathbb{P}(r_{k}|x_{1:k}, y_{1:k})$ is analytically tractable. Using
this analytical substructure, it is possible to implement an efficient
RBPF scheme which can reduce the variance of the estimation error. In
the sequel, we detail this RBPF framework for the MmNDS.

\section{A new RBPF for Markov modulated nonlinear state-space model}
\label{RBPF:desc}

In this section we outline a new RBPF framework exploiting the
conditionally finite state-space HMM.

\subsection{Description of the RBPF approach}
At time zero, the initial densities for the state and the regime
variables are respectively given by $p(x_0)$ and
$\mathbb{P}(r_0)\triangleq \mathbb{P}(r_{0}|x_{0})$, these can be
arbitrary but ear assumed to be known. We further assume favorable
mixing conditions as in \cite{CriDou:02}.

Suppose that we are at time $k-1$. We consider the extended target
density $p(r_{k-1},x_{0:k-1}|y_{1:k-1})$ which can be decomposed as
\begin{equation}
 \label{target_prev}
 p(r_{k-1},x_{0:k-1}|y_{1:k-1})
 = p(r_{k-1}|x_{0:k-1},y_{1:k-1})\,p(x_{0:k-1}|y_{1:k-1}) .
\end{equation}
The posterior propagation of the latent state $x_{k-1}$ can then be
targeted through a PF, where $p(x_{0:k-1}|y_{1:k-1})$ is represented
by a set of $N$ weighted random particles as
\begin{equation}
  \label{eq:mpf-NSSS}
  p(x_{0:k-1}|y_{1:k-1})
  \approx\sum_{i=1}^{N}w_{k-1}^{(i)}\delta(x_{0:k-1}-{x_{0:k-1}^{(i)}}).
\end{equation}
%with sample trajectories $x_{1:k-1}^{(i)}$ and weights $\omega_{k-1}^{(i)}$. Note that the marginal distribution $p(x_{k-1}|y_{1:k-1})$ can be obtained from \eqref{eq:mpf-NSSS} as
%\begin{equation}
%p(x_{k-1}|y_{1:k-1})\approx\sum_{i=1}^{N}\omega_{k-1}^{(i)}\delta(x_{k-1}-x_{k-1}^{(i)}).\label{eq:marginal-NSSS}
%\end{equation}

Now conditioned on $\{x_{0:k-1}, y_{1:k-1}\}$, the regime variable
$r_{k-1}$ follows a finite state-space HMM. As a result
$p(r_{k-1}|x_{0:k-1},y_{1:k-1})$ is analytically
tractable\footnote{Observe that this distribution depends on the PF
  path space representation $x_{0:k-1}$. It is well known that with
  time, particle filter suffers from a progressively impoverished
  particle representation. This is caused due to the effect of
  repeated resampling steps, leading to a path degeneracy problem
  \citep{CGM}. On the other hand, uniform convergence in time of the
  particle filter is known under the mixing assumptions as in
  \cite{CriDou:02}. This property ensures that any error is forgotten
  exponentially with time and can explain why the particle filter
  works for the marginal filter density.}, which is represented as
\begin{equation}
  \label{mode:prob}
  q_{k-1|k-1}^{(i)}(l)
  \triangleq \mathbb{P}(r_{k-1}=l|x_{0:k-1}^{(i)},y_{1:k-1}) ,
\end{equation}
for $l \in S$ and $i=1,\dots,N$. Now using \eqref{eq:mpf-NSSS} and
\eqref{mode:prob}, the extended target density in \eqref{target_prev}
can be represented as
\begin{equation}
  \label{cloud_prev}
  \left[x_{0:k-1}^{(i)}, w_{k-1}^{(i)}, \{q_{k-1|k-1}^{(i)}(l)\}_{l=1}^{s} \right]_{i=1}^{N}.
\end{equation}
Now having observed $y_k$, we want to propagate the extended target
density in \eqref{target_prev} to time $k$. This can be achieved in
the following steps (a)--(d):

\noindent\textbf{(a) Prediction step for conditional HMM filter:} this
is easily obtained as\footnote{Since each $q_{k-1|k-1}^{(i)}(\cdot)$
  is smaller than $1$, and the recursion involves multiplication by
  the terms, each less than $1$, some $q_{k|k-1}^{(i)}(\cdot)$ can
  become very small. For this numerical problem, it is better to work
  with $\log \big(q_{k|k-1}^{(i)}(\cdot)\big)$.}
\begin{subequations}
  \begin{align}
    q_{k|k-1}^{(i)}(l)
    &\triangleq \mathbb{P}(r_{k}=l|x_{0:k-1}^{(i)},y_{1:k-1}) \\ %
    \label{q_predt}
    &= \sum_{j=1}^{s}{\pi_{jl}}\ q_{k-1|k-1}^{(i)}(j), \quad (l,j) \in S .
  \end{align}
\end{subequations}

\noindent\textbf{(b) Prediction step for particle filter:}
at this stage, generate $N$ new samples $x_{k}^{(i)}$ from an
appropriate proposal kernel as
\begin{equation}
  \label{prop_kern}
  x_{k}^{(i)}
  \sim \pi(x_{k}|x_{0:k-1}^{(i)},y_{1:k}) .
\end{equation}
Then set $x_{0:k}^{(i)} = \{x_{0:k-1}^{(i)}, x_{k}^{(i)} \},$ for
$i=1,\dots,N$, representing the particle trajectories up to time $k$.

\noindent\textbf{(c) Update step for conditional HMM filter:}
noting that
\begin{multline}
  \mathbb{P}(r_{k}=l|x_{0:k},y_{1:k})\\                %
  \propto p(y_{k},x_{k}|r_{k}=l,x_{0:k-1},y_{1:k-1})\,
  \mathbb{P}(r_{k}=l|x_{0:k-1},y_{1:k-1}),
\end{multline}
we have
\begin{subequations}
  \begin{align}
    q_{k|k}^{(i)}(l)
    &\propto  p(y_{k},x_{k}^{(i)}|r_{k}=l,x_{0:k-1}^{(i)},y_{1:k-1})\,
    q_{k|k-1}^{(i)}(l)\\        %
    &\propto   p_{\theta_{l}}(y_{k}|x_{k}^{(i)},r_{k}=l)\,
    p_{\theta_{l}}(x_{k}^{(i)}|x_{k-1}^{(i)},r_{k}=l)\,
    q_{k|k-1}^{(i)}(l).
  \end{align}
\end{subequations}
Now defining
\begin{gather}
	\label{def:alpha}
  \alpha_{k}^{(i)}(l)
  \triangleq p_{\theta_{l}}(y_{k}|x_{k}^{(i)},r_{k}=l)\,
  p_{\theta_{l}}(x_{k}^{(i)}|x_{k-1}^{(i)},r_{k}=l)\,
  q_{k|k-1}^{(i)}(l)
\shortintertext{we obtain}
  \label{hmm:wt}
  q_{k|k}^{(i)}(l)
  = \frac{\alpha_{k}^{(i)}(l)}{\sum_{j=1}^{s}\alpha_{k}^{(i)}(j)},
\end{gather}
for $l\in S$ and $i=1,\dots,N$.

\noindent\textbf{(d) Update step for particle filter:}
as the continuous state can be recursively propagated using the
following relation:
\begin{equation}
  p(x_{0:k}|y_{1:k})
  \propto p(y_{k},x_{k}|x_{0:k-1},y_{1:k-1})\,p(x_{0:k-1}|y_{1:k-1}),
\end{equation}
the corresponding weight update equation for the particle filtering is
given by
\begin{subequations}
  \begin{align}
    \label{wt_upd}
    w_{k}^{(i)}
    &=\frac{p(x_{k}^{(i)}, y_{k} | x_{0:k-1}^{(i)}, y_{1:k-1})}
           {\pi_{k}(x_{k}^{(i)} |  x_{0:k-1}^{(i)}, y_{1:k})}
    \widetilde{w}_{k-1}^{(i)}\\ %
    \label{wt_norm}
    \widetilde{w}_{k}^{(i)}
    &= \frac{{w}_{k}^{(i)}}{\sum_{j=1}^{N}{w}_{k}^{(j)}} ,
  \end{align}
\end{subequations}
where $\{\widetilde{w}_{k}^{(i)}\}_{i=1}^{N}$ are the normalized
weights. The numerator\linebreak
$p(x_{k}^{(i)}, y_{k} | x_{0:k-1}^{(i)}, y_{1:k-1})$ can be obtained
as
%------------------------
%%\begin{align}
% \left(
%    \begin{pmatrix}
%
%    \end{pmatrix}
%%|\begin{array}{c} x_{k-1} \\ y_{k-1} \end{array}
%|\begin{pmatrix}
% x_{k-1} \\ y_{k-1}
%\end{pmatrix}
%\right)
%------------------------
%\begin{multline}
%p(x_{k}^{(i)}, y_{k} | x_{0:k-1}^{(i)}, y_{1:k-1}) = \\ \sum_{l=1}^{s} p(y_{k},x_{k}^{(i)}|r_{k}=l,x_{0:k-1}^{(i)},y_{1:k-1}) \ \mathbb{P}(r_{k}=l|x_{0:k-1}^{(i)},y_{1:k-1})\label{jt_lkd},
%\end{multline}
\begin{equation}
  \label{jt_lkd}
  p\Biggl({x_{k}^{(i)} \atop y_{k}} \Biggm|
  {x_{0:k-1}^{(i)} \atop y_{1:k-1}}\Biggr)
  = \sum_{l=1}^{s} p\Biggl({y_{k}\atop x_{k}^{(i)}}\Biggm|
  {r_{k}=l,x_{0:k-1}^{(i)}\atop y_{1:k-1}}\Biggr)\,
  \mathbb{P}\Biggl(r_{k}=l\Biggm|
  {x_{0:k-1}^{(i)}\atop y_{1:k-1}}\Biggr),
\end{equation}
which is basically given by the normalizing constant of
\eqref{hmm:wt}. Note that the marginal density $p(x_{k}|y_{1:k})$ can
be obtained as
\begin{equation}
  \label{eq:marginal-NSSS}
  p(x_{k}|y_{1:k})
  \approx\sum_{i=1}^{N}\widetilde{w}_{k}^{(i)}\delta(x_{k}-{x_{k}^{(i)}}).
\end{equation}

The posterior probability of the regime variable can now be obtained
as
\begin{subequations}
  \begin{align}
    \mathbb{P}(r_{k}=l|y_{0:k})
    &=  \int\mathbb{P}(r_{k}=l|x_{0:k},y_{0:k})
    p(x_{0:k}|y_{0:k})\,dx_{0:k} \\ %
    \label{reg_pos}
    &\approx\sum_{i=1}^{N} q_{k|k}^{(i)}(l)\widetilde{w}_{k}^{(i)} .
  \end{align}
\end{subequations}
The mean and variance of the marginal distribution in
\eqref{eq:marginal-NSSS} at time $k$ can be obtained from the weighted
particle representation as
\begin{subequations}
  \begin{align}
    \widehat{x}_k
    &= \sum_{i=1}^{N}\widetilde{w}_{k}^{(i)}{x_{k}^{(i)}}, \\ %
    \widehat{P}_k
    &= \sum_{i=1}^{N}\widetilde{w}_{k}^{(i)}
    (x_{k}^{(i)}-\widehat{x}_k)
    (x_{k}^{(i)}-\widehat{x}_k)^{T} ,
  \end{align}
\end{subequations}
where $(\cdot)^T$ denotes the transpose operation.  Let
$\widehat{m}^{(i)}_k$ and $\widehat{V}^{(i)}_k$ denote the mean and
variance of the conditional HMM filter. They are now obtained as
\begin{subequations}
  \begin{align}
    \label{mean:chmm}
    \widehat{m}^{(i)}_k
    &= \sum_{l=1}^{s}(r_{k}=l)\ q_{k|k}^{(i)}(l) ,\\ %
    \label{var:chmm}
    \widehat{V}^{(i)}_k
    &=
    \sum_{l=1}^{s}\{(r_{k}=l)-\widehat{m}^{(i)}_k\}
    \{(r_{k}=l)-\widehat{m}^{(i)}_k\}^{T}\,
    q_{k|k}^{(i)}(l) .
\end{align}
\end{subequations}
As noted earlier, the posterior of the regime variable is given by
\eqref{reg_pos}. Let $\widehat{m}_k$ and $\widehat{V}_k$ denote the
corresponding mean and variance, which can be obtained as
\begin{subequations}
  \begin{align}
    \label{reg_mean}
    \widehat{m}_k
    &= \sum_{i=1}^{N}\widetilde{w}_{k}^{(i)}\,\widehat{m}^{(i)}_{k},\\ %
    \label{reg_var}
    \widehat{V}_k &
    = \sum_{i=1}^{N}\widetilde{w}_{k}^{(i)}\,
    \Bigl[\widehat{V}^{(i)}_{k}+ (\widehat{m}^{(i)}_{k}-\widehat{m}_{k})
    (\widehat{m}^{(i)}_{k}-\widehat{m}_{k})^{T}\Bigr].
\end{align}
\end{subequations}

\begin{rem}
  For PF, a popular (but less efficient) choice for the proposal
  kernel is given by the state transition density $p(x_{k}|x_{k-1})$,
  which in this case can be obtained in the form of a weighted mixture
  density:
\begin{equation}
  p(x_{k}|x_{k-1}^{(i)})
  = \sum_{l=1}^{s}p_{\theta_{r_{k}}}(x_{k}|x_{k-1}^{(i)},r_{k}=l)\,q_{k|{k-1}}^{(i)}(l),
\end{equation}
\end{rem}
where $p_{\theta_{r_{k}}}(x_{k}|x_{k-1}^{(i)},r_{k}=l)$ is specified
in \eqref{reg_dyn}.
%
%
%
%%\begin{align}
%\label{jt_xe}
% \left(
%    \begin{pmatrix}
%      x_{k}\\ e_k
%    \end{pmatrix}
%%|\begin{array}{c} x_{k-1} \\ y_{k-1} \end{array}
%|\begin{pmatrix}
% x_{k-1} \\ y_{k-1}
%\end{pmatrix}
%\right)

\subsection{Algorithmic summary}

The new RBPF for the MmNDS is summarized in Algorithm~\ref{alg:rbpf}.

\begin{algorithm}
  \caption{RBPF for MmNDM}\label{alg:rbpf}

  \underline{Initialization:}\\
  For each particle $i=1,\dots,N$ do
  \begin{itemize}[noitemsep,topsep=0pt,parsep=0pt,partopsep=0pt]
  \item Sample $x_{0}^{(i)}\sim p(x_{0})$,
  \item Set initial weights $w_{0}^{(i)}=\frac{1}{N}$,
  \item Set initial $q_{0|0}^{(i)}(l) \triangleq \mathbb{P}(r_{0}=l|x_{0}^{(i)})$, \ \ $l=1,\dots,s$
  \end{itemize}

  \underline{Iterations:}\\
  Set the resampling threshold $\eta$; \\
  For $k=1,2,\ldots$ do
  \begin{itemize}[noitemsep,topsep=0pt,parsep=0pt,partopsep=0pt]
  \item For each particle $i=1,\dots,N$ do
    \begin{itemize}[noitemsep,topsep=0pt,parsep=0pt,partopsep=0pt]
    \item Compute $q_{k|{k-1}}^{(i)}(l)$ using \eqref{q_predt}
    \item Sample $x_{k}^{(i)}\sim \pi(x_{k|\cdot})$ using
      \eqref{prop_kern}
    \item Set $x_{0:k}^{(i)} \triangleq (x_{0:k-1}^{(i)},
      x_{k}^{(i)})$
    \item Compute $\alpha_{k}^{(i)}(l)$ using \eqref{def:alpha}
    \item Compute $q_{k|{k}}^{(i)}(l)$ using \eqref{hmm:wt}
    \item Compute $w_{k}^{(i)}$ using \eqref{wt_upd} and
      \eqref{jt_lkd} as
      \begin{equation*}
        w_{k}^{(i)}
        = \frac{\sum_{j=1}^{s}\alpha_{k}^{(i)}(j)}
               {\pi_{k}(x_{k}^{(i)} |  x_{1:k-1}^{(i)}, y_{1:k})}
               \widetilde{w}_{k-1}^{(i)}
      \end{equation*}
    \end{itemize}
  \item Normalize the weights using  \eqref{wt_norm}
  \item Compute $N_{\text{eff}}=\frac{1}{\sum_{i=1}^{N}(\widetilde{w}_{k}^{(i)})^{2}}$.
    \begin{itemize}[noitemsep,topsep=0pt,parsep=0pt,partopsep=0pt]
    \item If $N_{\text{eff}}\leq \eta$, resample the particles. Let the
      resampled particles be $i^{\ast}=1,\dots,N$.
    \item Copy the corresponding $q_{k|k}^{(i^{\ast})}(l)$ and set
      $\widetilde{w}_{k}^{(i^{\ast})}=\frac{1}{N}$.
    \end{itemize}
  \end{itemize}
\end{algorithm}

\section{Relation to other similar models}
\label{RBPF_comp}

Here we compare our RBPF model to other existing models exploiting
similar conditional substructure. Similar conditionally finite
state-space HMM have earlier been considered by \cite{Doucet2000a} as
well as\linebreak%
 \cite{Andrieu02}, although, each framework is fundamentally
different. The differences are emphasized below.

In our case $({x_k, y_k})$ follows a nonlinear state-space model,
which is modulated by a finite state hidden Markov process
${r_k}$. Hierarchically $r_k$ is at the top level and is not
influenced by $x_k$. This is different from the hierarchical
conditionally finite state-space HMM in \cite{Doucet2000a}, where
$({r_k, y_k})$ follows a finite state-space hidden Markov process,
which is modulated by a another (hidden) Markov process $c_k$. Here
$c_k$ is at the top of hierarchy and is not influenced by $r_k$. In
contrast, \cite{Andrieu02} considered a partially observable finite
state-space HMM, where ${r_k}$ is a finite state hidden Markov
process, $y_k$ is a latent data process and $z_k$ is observed data
process. Conditioned on the sequence $z_{1:k}$, here $({r_k, y_k})$
follows a finite state-space HMM.

\section{Concluding remarks}\label{conclusion}

Markov modulated nonlinear state-space model, although less well
explored, appears quite naturally in many applications of
interest. The model implies that the system can switch between
different nonlinear dynamic regimes. The regime state is governed by a
regime variable, which follows a homogeneous finite state first-order
Markov process. In this article, the associated online inference
problem for such model is addressed. In particular, a new RBPF is
proposed for such inference tasks. This RBPF scheme exploits the
analytical marginalization of the regime variable using the
conditional HMM structure. This results in improved performance over a
standard particle filter in terms of variance of the estimation
error. Moreover for a standard particle filter where the regime state
is also represented by the particles, degeneracy is commonly observed
around regime transition \citep{Hans05}. In our RBPF implementation, as
the regime variable follows a conditionally analytical substructure,
hence the degeneracy is expected to be less severe.

\section*{Acknowledgment}

The authors would like to thank COOP-LOC, funded by SSF and CADICS,
funded by Swedish Research Council (VR) for the financial supports.


\appendix

\section{Sketch of the variance reduction mechanism through
  Rao-Blackwellization}
\label{apndx}

Let $\theta$ be an unknown parameter and $Y$ be the random variable
corresponding to the observed data. Let $\widehat{\theta}(Y)$ be any
kind of estimator of $\theta$. Further, if $T$ be the sufficient
statistics for $Y$, then the Rao-Blackwell theorem states that the
following estimator
\begin{equation}
  % \widehat{\theta}_{RB}(T(Y)) = \mathbb{E}[\widehat{\theta}(Y)|T(Y) \]
	\label{RB_est}
  \widehat{\theta}_{RB}(T)
  = \mathbb{E}[ \widehat{\theta}(Y)|T]
\end{equation}
is typically a better estimator of $\theta$, and is never worse. The
transformed estimator $\widehat{\theta}_{RB}(T)$ using the sufficient
statistics is known as the Rao-Blackwell estimator
\citep{lehmann:1983}.

Now suppose $X$ is a random variable admitting a probability density
function $p(x)$. Further, let $g(\cdot)$ be a function of $X$ and
$\Phi$ be a test function given as the expectation of $g(X)$
\begin{gather}
  \label{def_phi}
  \Phi
  = \mathbb{E}	[g(X)]= \int g(x)p(x)\,dx . %
  \intertext{A Monte Carlo based estimator of $\Phi$ can be obtained
    as} %
  \label{phi_MC}
  \widehat{\Phi}_{MC}(X)
	= \frac{1}{N}\sum_{i=1}^{N}\ g(x^{(i)}), %
  \intertext{where $x^{(i)},\ i=1,\dots,N$ are generated according to
    $p(x)$. The variance of this estimator is} %
  \Var\bigl(\widehat{\Phi}_{MC}(X)\bigr)
	=\frac{\Var[g(X)]}{N},
\end{gather}
provided that the variance of $g(X)$ is finite.

Now suppose that $X$ is a random vector which can be split into two
components as $X = (\Xi, \Lambda)^T$. Using \eqref{phi_MC}, we have
\begin{gather}
  \label{phi_MC_full}
  \widehat{\Phi}_{MC}(\Xi, \Lambda)
	= \frac{1}{N}\sum_{i=1}^{N}\ g(\xi^{(i)}, \ \lambda^{(i)}). %
  \intertext{Using \eqref{def_phi} and law of iterated expectations,
    we can write}               %
  \label{it_exp}
  \Phi
  = \mathbb{E}\Big[\mathbb{E}\{g(\Xi, \Lambda)|\Xi\}\Big]. %
  \intertext{We can subsequently define the following Rao-Blackwell
    estimator using \eqref{RB_est} and \eqref{it_exp} as} %
	\widehat{\Phi}_{RB}(\Xi)
  =\mathbb{E}\Bigl[\widehat{\Phi}_{MC}(\Xi, \Lambda)\Bigm| \Xi\Bigr].
\end{gather}

Now using the law of total variance
\begin{gather}
	\Var(\Phi)
  = \Var\Big(\mathbb{E}[\Phi|\Xi]\Big)+
  \underbrace{\mathbb{E}\Big(\Var[\Phi|\Xi]\Big)}_{\ge 0}. %
  \shortintertext{Consequently, we have} %
  \Var\Big(\widehat{\Phi}_{MC}(\Xi, \lambda)\Big)
  \ge \Var\Big(\widehat{\Phi}_{RB}(\Xi)\Big) .
\end{gather}
% The last step shows the gain in variance reduction due to the
% Rao-Blackwellization.
Rao-Blackwellization is useful when $\mathbb{E}[\Phi|\Xi]$ can be
computed efficiently. This happens \eg, when part of the integration
in \eqref{def_phi} is analytically tractable.

\bibliographystyle{model2-names}
\bibliography{IEEEfull,bibtex_db}

\begin{thebibliography}{19}
\expandafter\ifx\csname natexlab\endcsname\relax\def\natexlab#1{#1}\fi
\providecommand{\url}[1]{\texttt{#1}}
\providecommand{\href}[2]{#2}
\providecommand{\path}[1]{#1}
\providecommand{\DOIprefix}{doi:}
\providecommand{\ArXivprefix}{arXiv:}
\providecommand{\URLprefix}{URL: }
\providecommand{\Pubmedprefix}{pmid:}
\providecommand{\doi}[1]{\href{http://dx.doi.org/#1}{\path{#1}}}
\providecommand{\Pubmed}[1]{\href{pmid:#1}{\path{#1}}}
\providecommand{\bibinfo}[2]{#2}
\ifx\xfnm\relax \def\xfnm[#1]{\unskip,\space#1}\fi
%Type = Article
\bibitem[{Andrieu et~al.(2003)Andrieu, Davy and Doucet}]{AndrieuMDAD03}
\bibinfo{author}{Andrieu, C.}, \bibinfo{author}{Davy, M.},
  \bibinfo{author}{Doucet, A.}, \bibinfo{year}{2003}.
\newblock \bibinfo{title}{{Efficient particle filtering for jump Markov
  systems}}.
\newblock \bibinfo{journal}{{IEEE} Transactions on Signal Processing}
  \bibinfo{volume}{51}, \bibinfo{pages}{1762--1770}.
%Type = Article
\bibitem[{Andrieu and Doucet(2002)}]{Andrieu02}
\bibinfo{author}{Andrieu, C.}, \bibinfo{author}{Doucet, A.},
  \bibinfo{year}{2002}.
\newblock \bibinfo{title}{Particle filtering for partially observed {G}aussian
  state space models}.
\newblock \bibinfo{journal}{Journal of the Royal Statistical Society Series B}
  \bibinfo{volume}{64}, \bibinfo{pages}{827--836}.
%Type = Book
\bibitem[{Barber(2012)}]{barberBRML2012}
\bibinfo{author}{Barber, D.}, \bibinfo{year}{2012}.
\newblock \bibinfo{title}{{B}ayesian Reasoning and Machine Learning}.
\newblock \bibinfo{publisher}{Cambridge University Press}.
%Type = Article
\bibitem[{Capp{\'e} et~al.(2007)Capp{\'e}, Godsill and Moulines}]{CGM}
\bibinfo{author}{Capp{\'e}, O.}, \bibinfo{author}{Godsill, S.J.},
  \bibinfo{author}{Moulines, E.}, \bibinfo{year}{2007}.
\newblock \bibinfo{title}{An overview of existing methods and recent advances
  in sequential \uppercase{M}onte \uppercase{C}arlo}.
\newblock \bibinfo{journal}{Proceedings of the {IEEE}} \bibinfo{volume}{95
  (5)}, \bibinfo{pages}{899--924}.
%Type = Article
\bibitem[{Carvalho and Lopes(2003)}]{carvalhoHL07}
\bibinfo{author}{Carvalho, C.M.}, \bibinfo{author}{Lopes, H.F.},
  \bibinfo{year}{2003}.
\newblock \bibinfo{title}{Simulation-based sequential analysis of {M}arkov
  switching stochastic volatility models}.
\newblock \bibinfo{journal}{Computational Statistics \& Data Analysis}
  \bibinfo{volume}{51}, \bibinfo{pages}{4526--4542}.
%Type = Article
\bibitem[{Chen and Liu(2000)}]{chen:liu:00:mixture}
\bibinfo{author}{Chen, R.}, \bibinfo{author}{Liu, J.}, \bibinfo{year}{2000}.
\newblock \bibinfo{title}{Mixture \uppercase{K}alman filters}.
\newblock \bibinfo{journal}{Journal of the Royal Statistical. Society Series B}
  \bibinfo{volume}{62}, \bibinfo{pages}{493--508}.
%Type = Article
\bibitem[{Chopin(2007)}]{Chopin04}
\bibinfo{author}{Chopin, N.}, \bibinfo{year}{2007}.
\newblock \bibinfo{title}{Central limit theorem for sequential {M}onte {C}arlo
  methods and its application to {B}ayesian inference}.
\newblock \bibinfo{journal}{The Annals of Statistics} \bibinfo{volume}{32},
  \bibinfo{pages}{2385--2411}.
%Type = Article
\bibitem[{Crisan and Doucet(2002)}]{CriDou:02}
\bibinfo{author}{Crisan, D.}, \bibinfo{author}{Doucet, A.},
  \bibinfo{year}{2002}.
\newblock \bibinfo{title}{A survey of convergence results on particle filtering
  methods for practitioners}.
\newblock \bibinfo{journal}{{IEEE} Transactions on Signal Processing}
  \bibinfo{volume}{50}, \bibinfo{pages}{736--746}.
%Type = Article
\bibitem[{Doucet et~al.(2000)Doucet, Godsill and Andrieu}]{Doucet2000a}
\bibinfo{author}{Doucet, A.}, \bibinfo{author}{Godsill, S.},
  \bibinfo{author}{Andrieu, C.}, \bibinfo{year}{2000}.
\newblock \bibinfo{title}{On sequential \uppercase{M}onte \uppercase{C}arlo
  sampling methods for \uppercase{B}ayesian filtering}.
\newblock \bibinfo{journal}{Statistics and Computing} \bibinfo{volume}{10},
  \bibinfo{pages}{197--208}.
%Type = Inproceedings
\bibitem[{Doucet and Johansen(2011)}]{Doucet:Johansen11}
\bibinfo{author}{Doucet, A.}, \bibinfo{author}{Johansen, A.M.},
  \bibinfo{year}{2011}.
\newblock \bibinfo{title}{A tutorial on particle filtering and smoothing:
  Fifteen years later}, in: \bibinfo{booktitle}{Oxford Handbook of Nonlinear
  Filtering}, \bibinfo{publisher}{Oxford University Press}.
%Type = Article
\bibitem[{Driessen and Boers(2005)}]{Hans05}
\bibinfo{author}{Driessen, H.}, \bibinfo{author}{Boers, Y.},
  \bibinfo{year}{2005}.
\newblock \bibinfo{title}{Efficient particle filter for jump {M}arkov nonlinear
  systems}.
\newblock \bibinfo{journal}{IEE Proceedings- radar, sonar and navigation}
  \bibinfo{volume}{152}, \bibinfo{pages}{323--326}.
%Type = Book
\bibitem[{Fruhwirth-Schnatter(2007)}]{Fruhwirth07}
\bibinfo{author}{Fruhwirth-Schnatter, S.}, \bibinfo{year}{2007}.
\newblock \bibinfo{title}{Finite Mixture and Markov Switching Models}.
\newblock \bibinfo{publisher}{Springer}.
%Type = Article
\bibitem[{Hendeby et~al.(2010)Hendeby, Karlsson and Gustafsson}]{Hendeby10}
\bibinfo{author}{Hendeby, G.}, \bibinfo{author}{Karlsson, R.},
  \bibinfo{author}{Gustafsson, F.}, \bibinfo{year}{2010}.
\newblock \bibinfo{title}{The {R}ao-{B}lackwellized particle filter: A filter
  bank implementation}.
\newblock \bibinfo{journal}{EURASIP Journal on Advances in Signal Processing}
  \bibinfo{volume}{2010}.
%Type = Book
\bibitem[{Lehmann(1983)}]{lehmann:1983}
\bibinfo{author}{Lehmann, E.L.}, \bibinfo{year}{1983}.
\newblock \bibinfo{title}{Theory of Point Estimation}.
\newblock Probability and Mathematical Statistics, \bibinfo{publisher}{John
  Wiley \& Sons, Ltd}.
%Type = Article
\bibitem[{Luo and Mao(2007)}]{Luo07}
\bibinfo{author}{Luo, Q.}, \bibinfo{author}{Mao, X.}, \bibinfo{year}{2007}.
\newblock \bibinfo{title}{Stochastic population dynamics under regime
  switching}.
\newblock \bibinfo{journal}{Journal of Mathematical Analysis and Applications,}
  \bibinfo{volume}{334}, \bibinfo{pages}{69--84}.
%Type = Book
\bibitem[{Mamon and Elliott(2007)}]{mamon2007hidden}
\bibinfo{author}{Mamon, R.S.}, \bibinfo{author}{Elliott, R.J.},
  \bibinfo{year}{2007}.
\newblock \bibinfo{title}{Hidden Markov models in finance}.
\newblock \bibinfo{publisher}{Springer, New York}.
%Type = Book
\bibitem[{Ristic et~al.(2004)Ristic, Arulampalam and Gordon}]{Risticbook04}
\bibinfo{author}{Ristic, B.}, \bibinfo{author}{Arulampalam, S.},
  \bibinfo{author}{Gordon, N.}, \bibinfo{year}{2004}.
\newblock \bibinfo{title}{Beyond the Kalman Filter: Particle Filters for
  Tracking Applications}.
\newblock \bibinfo{publisher}{Artech House}.
%Type = Inproceedings
\bibitem[{Saha et~al.(2010)Saha, Ozkan, Gustafsson and Smidl}]{SahaOGS09}
\bibinfo{author}{Saha, S.}, \bibinfo{author}{Ozkan, E.},
  \bibinfo{author}{Gustafsson, F.}, \bibinfo{author}{Smidl, V.},
  \bibinfo{year}{2010}.
\newblock \bibinfo{title}{Marginalized particle filters for {B}ayesian
  estimation of {G}aussian noise parameters.}, in:
  \bibinfo{booktitle}{Proceedings of 13th International Conference on
  Information Fusion (FUSION)}.
%Type = Article
\bibitem[{Sch{\"o}n et~al.(2005)Sch{\"o}n, Gustafsson and Nordlund}]{Schon05}
\bibinfo{author}{Sch{\"o}n, T.}, \bibinfo{author}{Gustafsson, F.},
  \bibinfo{author}{Nordlund, P.J.}, \bibinfo{year}{2005}.
\newblock \bibinfo{title}{Marginalized particle filter for mixed
  linear/nonlinear state space models}.
\newblock \bibinfo{journal}{IEEE Transaction on Signal Processing}
  \bibinfo{volume}{53}, \bibinfo{pages}{2279--2289}.

\end{thebibliography}

%% not want to use model2-names.bst.



%% \bibitem must have one of the following forms:
%%   \bibitem[Jones et al.(1990)]{key}...
%%   \bibitem[Jones et al.(1990)Jones, Baker, and Williams]{key}...
%%   \bibitem[Jones et al., 1990]{key}...
%%   \bibitem[\protect\citeauthoryear{Jones, Baker, and Williams}{Jones
%%       et al.}{1990}]{key}...
%%   \bibitem[\protect\citeauthoryear{Jones et al.}{1990}]{key}...
%%   \bibitem[\protect\astroncite{Jones et al.}{1990}]{key}...
%%   \bibitem[\protect\citename{Jones et al., }1990]{key}...
%%   \harvarditem[Jones et al.]{Jones, Baker, and Williams}{1990}{key}...
%%


% \end{thebibliography}

\end{document}